\documentclass[12pt]{article}
\usepackage[dvipdfmx]{graphicx}
\usepackage{hyperref,amssymb,amsmath}
\usepackage{bbold}

\usepackage{mathtools}
\usepackage{bm}
\usepackage{url}
\usepackage{color}

\newcommand{\LQ}{\Lambda_{\rm QCD}}
\newcommand{\alfs}{\alpha_{s}}
\newcommand{\msbar}{\overline{\rm MS}}

\newcommand{\bea}{\begin{eqnarray}}
\newcommand{\eea}{\end{eqnarray}}
\newcommand{\simgt}{\hbox{ \raise3pt\hbox to 0pt{$>$}\raise-3pt\hbox{$\sim$} }}
\newcommand{\simlt}{\hbox{ \raise3pt\hbox to 0pt{$<$}\raise-3pt\hbox{$\sim$} }}

\newcommand{\be}{\begin{equation}}
\newcommand{\ee}{\end{equation}}

\newcommand{\non}{\nonumber \\}
\begin{document}

\begin{titlepage}

    \begin{flushright}
      \normalsize TU--1142\\
      \today
    \end{flushright}

\vskip2.5cm
\begin{center}
\Large
Renormalon subtraction using Fourier transform:\\
Analyses of simplified models
\end{center}

\vspace*{0.2cm}
\begin{center}
{\sc Yuuki Hayashi} \\[5mm]
  {\small\it Department of Physics, Tohoku University}\\[0.1cm]
  {\small\it Sendai, 980-8578 Japan}

\end{center}

\vspace*{2.8cm}
\begin{abstract}
\small
For precise QCD prediction of observables, the ambiguity due to renormalons in perturbative calculations should be appropriately separated from Wilson coefficients in the framework of the operator-product-expansion. Recently, a new method has been developed which utilizes the properties of Fourier transform to separate multiple renormalons simultaneously from the Wilson coefficients. To understand how this method works analytically, we perform a renormalon separation from various toy models with the one-loop beta function. We confirm that each of the results is consistent with the theoretical expectations. In addition, we present a new formula for the resummation of UV renormalons and study its validity using one of the toy models.
\end{abstract}


\vfil
\end{titlepage}






\section{Introduction}
The development of computational technology in recent decades
has enabled calculation of higher-order corrections in perturbative QCD for many observables.
As higher-order calculations have been achieved,
it has become clear that a proper treatment of low-energy non-perturbative QCD effects
is necessary to make even more precise theoretical predictions. 
The accuracy of theoretical predictions from perturbation theory 
is limited by renormalons, 
and from considerations based on the operator-product-expansion (OPE) framework, 
theoretical ambiguities induced from renormalons mix with indeterminacies of non-perturbative effects. 
It is not possible to treat only non-perturbative effects neglecting renormalon effects, 
and we need a method to separate the contribution of renormalons from perturbative calculations.
Recent studies on this direction are discussed in refs.~\cite{Takaura:2018lpw,Ayala:2019uaw,Ayala:2019hkn,Ayala:2020odx,Takaura:2020byt}.

In refs.~\cite{Hayashi:2020ylq,Hayashi:2021vdq}, of which the present author is one of the authors,
a method has been developed for separation of renormalons 
from an observable with a single hard scale $Q$ by using the Fourier transform (FTRS method). 
This is an extension of the method applied to the QCD potential \cite{Sumino:2005cq}. 
In this method, we first perform a Fourier transform of a divergent series 
to generate a convergent series in the virtual ``momentum'' space.
Renormalons are evaluated as the divergence of inverse Fourier transform,
which stems from the singularity in the momentum space.
Then the renormalons are subtracted
by removing the divergence of the one-dimensional momentum integral
by taking the principal value (PV). 
The subtracted renormalons are evaluated as 
a closed-path integral surrounding the singularity in the momentum space.
They take a consistent form in the OPE
and can be absorbed into the non-perturbative matrix elements of the OPE. 
It is shown that the result obtained by this method is equivalent to 
the result of the well-known PV prescription, which is defined by using the principal value of the Borel integral.
One of the unique features of this method is that
it enables us to obtain the result of the PV prescription more precisely,
in a systematic manner,
as higher corrections in the momentum space are incorporated.
Moreover, the multiple renormalons can be subtracted simultaneously
by proper adjustment of the parameters in the Fourier transform.
All the results of the test analyses obtained so far are consistent with theoretical expectations.

We would like to understand these properties more explicitly and analytically 
for future applications of the FTRS method to a wide range of observables.
In this paper, in order to investigate how the method works, 
we use simplified toy models for demonstrations.
We perform analyses with the one-loop beta function
so that the computation can be treated as analytically as possible to show clearly what happens.
All of the results of the FTRS method are obtained by truncating the momentum-space series
and compared to the results of the PV prescription.
The toy models considered in this papar are
(A) the QCD running coupling constant $\alfs(Q)$,
(B) a model which contains IR renormalons at $u=1,\,2$ in the Borel plane,
(C) a model which contains an IR renormalon at $u=1$
and a UV renormalon at $u=-1$.
These analyses answer each of the following questions: 
(Q-A) What happens to a convergent series if the FTRS method is applied to it? 
(Q-B) How is the effect of IR renormalons suppressed in the perturbative analysis? 
(Q-C) Is the FTRS method useful for the perturbative series containing UV renormalons?

Compared to the study of ref.~\cite{Hayashi:2021vdq}, 
the following issues on the treatment of the perturbative series have been improved:
(i) the formula for resummation of all the artificial UV renormalons 
(ii) a new formula for the resummation of the UV renormalons which are included
originally in the leading Wilson coefficient.

\vspace{5mm}
This paper is organized as follows.
In Sec.~\ref{sec2}, we briefly review our method for renormalon subtraction using Fourier transform.
It contains a formula for the resummation of the generated UV renormalons in the momentum space.
This is a modified version of the formula in App.~E of ref.~\cite{Hayashi:2021vdq}.
In Sec.~\ref{sec3}, we demonstrate the renormalon subtraction using our method in the analyses of simplified toy models.
In Sec.~\ref{sec3-1}, we apply the method to $\alfs(Q)$ 
to investigate what the method causes to a convergent series.
In Sec.~\ref{sec3-2}, we study the subtraction of the IR renormalons at $u=1$ and $u=2$
in the analysis of a model constructed from the corresponding renormalon poles in the Borel plane.
In Sec.~\ref{sec3-3}, the resummation of UV renormalons in the original series is investigated.
We use the toy model with the IR renormalon at $u=1$ and the UV renormalon at $u=-1$.
Sec.~\ref{sec4} is devoted to the conclusions and discussions.

\section{Brief review of FTRS method}
\label{sec2}
\subsubsection*{Renormalon ambiguity in the context of OPE and RGE}
In this paper, all the discussions are given with one-loop beta function,
which simplifies the calculation of the FTRS method.
For the N$^k$LL case in the momentum space (which includes beyond one-loop beta function), 
see Sec. 2 of ref.~\cite{Hayashi:2021vdq}. 

Let us consider the operator-product-expansion (OPE) of a 
dimensionless observable ${\rm Obs}(Q)$ with a hard scale $Q(\gg \LQ)$ given by
\be
{\rm Obs}(Q)=C_0(Q)+C_1(Q)\frac{\langle O_1\rangle}{Q^{2u_1}}
+C_2(Q)\frac{\langle O_2\rangle}{Q^{2u_2}}+\cdots.
\ee
$C_i(Q)$ $(i=0,1,2,...)$ denotes the Wilson coefficient computed from the UV degrees of freedom.
$O_i$ $(i=1,2,...)$ is the operator in the low energy effective field theory
with dimension $2u_i$.
In QCD, the perturbation theory can be used to calculate $C_i(Q)$ due to the asymptotic freedom.
The perturbative expansion of the leading Wilson coefficient $C_0$ is given by
\be
\big[C_0(Q)\big]_{\rm PT}=\sum_{n=0}^\infty c_n(L_Q)\alpha_s^{n+1}
=\sum_{n=0}^\infty c_n(0)\alpha_s(Q)^{n+1}\,,
\label{XQ-PT}
\ee
where $\alfs=\alfs(\mu)$ and $L_Q=\log\big(\mu^2/Q^2)$.
In the following, 
$\alfs$ without the argument always stands for $\alfs(\mu)$.
$c_n(L_Q)$ is a polynomial of $L_Q$ obtained 
from the following relation,
\bea
\sum_{n=0}^\infty c_n(L_Q)\alpha_s^{n+1}
&=&e^{L_Q\hat{H}}\sum_{n=0}^\infty c_n(0)\alpha_s^{n+1}\non
&=&\sum_{m=0}^\infty\frac{1}{m!}L_Q^m\hat{H}^m\,\sum_{n=0}^\infty c_n(0)\alpha_s^{n+1}\,,
\label{RGinv}
\eea
where $\hat{H}=-\beta(\alpha_s)\big(\partial/\partial \alpha_s\big)$.
$\beta(\alpha_s)$ is the one-loop QCD beta function given by
\be
\beta(\alpha_s)=-b_0\alpha_s^2\,,\quad
b_0=\frac{1}{4\pi}\bigg(11-\frac{2}{3}n_f\bigg)\,.
\ee

The Wilson coefficient can also be represented as an integral (the Borel integral) given by
\be
\big[C_0(Q)\big]_{\rm Borel}=\frac{1}{b_0}\int_0^\infty \!\!\!\!du\,e^{-\frac{u}{b_0\alpha_s}}B_{C_0}(u)\,,
\label{Borel-def}
\ee
where the Borel transform $B_{C_0}(u)$ is given by
\be
B_{C_0}(u)=\sum_{n=0}^\infty\frac{c_n(L_Q)}{n!}\big(u/b_0\big)^n\,.
\ee 

It is conjectured that if $C_0(Q)$ contains IR renormalons, 
$\big[C_0(Q)\big]_{\rm PT}$
diverges as $c_n(0)\sim n!$ at high orders.
Correspondingly, 
the definition of $\big[C_0(Q)\big]_{\rm Borel}$ becomes ill-defined
since there emerge singularities on the positive real axis in the complex Borel $(u)$ plane.
To construct the perturbative calculation in the presence of renormalons,
the contour-deformed Borel integral is considered:
\be
\big[C_0(Q)\big]_{\rm Borel}^{\pm}=\frac{1}{b_0}\int_{C_\pm} \!\!\!\!du\,e^{-\frac{u}{b_0\alpha_s}}B_{C_0}(u)\,.
\label{Borelpm}
\ee 
In the complex $u$ plane
the integral contours $C_\pm(u)$ connect $0\pm i\epsilon$ and $+\infty\pm i\epsilon$
infinitesimally above/below the positive real axis on which the pole singularities are located.
This integral gives a finite value but
contains the ambiguous part dependent on the contour ($C_+$ or $C_-$).
The size of the theoretical ambiguity $\delta C_0$ induced by renormalons is defined 
from the corresponding singularities in the Borel $(u)$ plane,
\be
\delta C_0(Q)=\frac{1}{2ib_0}\int_{C_+-C_-} \!\!\!\!\!\!\!\!\!\!du\,e^{-\frac{u}{b_0\alpha_s}}B_{C_0}(u)\,.
\label{deltaX}
\ee
Then eq.~\eqref{Borelpm} is separated into two parts:
\be
\big[C_0(Q)\big]_{\rm Borel}^{\pm}=\big[C_0(Q)\big]_{\rm PV}\pm i\,\delta C_0(Q)\,,
\ee
where $\big[C_0(Q)\big]_{\rm PV}$ is given by
\be
\big[C_0(Q)\big]_{\rm PV}=\frac{1}{b_0}\int_{0,\rm PV}^\infty\!du\,e^{-\frac{u}{b_0\alpha_s}}B_{C_0}(u)\,.
\label{X-PV}
\ee
This renormalon separation procedure is called ``PV prescription''.
The PV prescription gives the picture that contributions from IR renormalons are minimally subtracted 
from the Wilson coefficient $C_0(Q)$.
The FTRS method gives a way to calculate eq.~\eqref{X-PV} systematically
from a finite number of terms in the perturbative expansion (see the next section).

We assume that the ambiguous part [$\pm i\,\delta C_0$] induced from renormalons 
are absorbed into non-perturbative terms in OPE 
and canceled against the same size ambiguities of the non-perturbative matrix element \cite{Mueller:1984vh}.
Then the location of a (IR) renormalon singularity in the complex $u$ plane
and the form of $\delta C_0$ due to this singularity can be determined,
up to an overall normalization,
by OPE and RGE from the corresponding non-perturbative matrix element.
For instance, the leading renormalon of $C_0$, which is located at $u=u_1$,
causes an ambiguous part which scales as $1/Q^{2u_1}$.
It is canceled with the leading non-perturbative matrix element 
$\langle O_1\rangle$ with dimension $2u_1$.\footnote{
Conversely, from the set of possible non-perturbative operators, 
we can find the locations of possible renormalons that $C_0$ can contain.
}
The form of $\delta C_0$ from the singularity at $u=u_1$ is given by
\be
\delta C_0(Q)\big|_{u=u_1}\propto\left(\frac{\LQ^2}{Q^2}\right)^{\! u_1}\!\!\!
\alpha_s(Q)^{\gamma_0/b_0}\,,
\label{renu}
\ee
where $\gamma_0$ denotes the one-loop coefficient of the anomalous dimension of operator $O_1$,
\be
\bigg[\mu^2\frac{d}{d\mu^2}+\gamma(\alpha_s)\bigg]O_1=0\,,\quad
\gamma(\alfs)=\gamma_0\alpha_s\,.
\ee
$\LQ$ denotes the dynamical scale of QCD at the one-loop level, 
which is given by
${\LQ^2/\mu^2}=e^{-\frac{1}{b_0\alpha_s}}$.
Beyond the one-loop approximation, 
the high order term in eq.~\eqref{renu} is fully determined by combination of 
the high order corrections of the beta function, anomalous dimension, and the Wilson coefficient $C_1$ \cite{Beneke:1998ui}.

\subsubsection*{Renormalon subtraction using Fourier transform}
We explain the new method called ``FTRS method'' to compute the renormalon-subtracted Wilson coefficient using the Fourier transform:
\be
\big[C_0(Q)\big]_{\rm FTRS}=\frac{r^{-2au'-1}}{2\pi^2}\int_{0,\rm PV}^\infty \!\!d\tau\,\tau\sin\big(\tau r\big)\tilde{C}_0(\tau)\,,
\label{FTRSdef}
\ee
where $a,\,u'$ are the parameters of this method,
and $r=Q^{-1/a}$.
$\tilde{C}_0(\tau)$ represents the Fourier transform of $C_0(Q)$,
from the $r(=Q^{-1/a})$ space to the momentum $\tau$ space,
given by
\be
\tilde{C}_0(\tau)=\int d^3\vec{x}\,e^{-i\vec{\tau}\cdot\vec{x}}r^{2au'}C(r^{-a})\quad;|\vec{x}|=r\,,
\label{tildeX}
\ee
$\tilde{C}_0(\tau)$ contains the singularities
on the real axis in the complex $\tau$ plane.
In fact, this method gives a result equivalent to the PV prescription result [eq.\eqref{X-PV}].
A proof of this statement is given in App. C of ref.~\cite{Hayashi:2021vdq}
under reasonable assumptions.

Our method requires a proper definition of the Fourier transform.
For simplicity of analysis, we make the following assumptions in this paper:
(i)The Wilson coefficient of non-perturbative matrix element is exactly one,
and there is no anomalous dimension for the corresponding operator $O_1$.
(ii)Singularities on the positive real axis in the Borel plane are 
simple poles.\footnote{
 It is shown that the FTRS method can be applied beyond the assumptions (i) and (ii).
See App.~B and C of ref.~\cite{Hayashi:2021vdq} 
}
(iii)There are no singularities except 
those which we suppress by the sine factor (on the 
positive real axis) in the Borel plane in the momentum space.
Then the form of $\delta C_0(Q)$ from the $u=u_*$ renormalon is given by
\be
\delta C_0(Q)\big|_{u=u_*}=A_{u_*}\left(\frac{\LQ^2}{Q^2}\right)^{\! u_*}\,,
\ee
where $A_{u_*}$ represents the normalization of the renormalon.
The value of $u_*$ is associated with the dimension $d$, of which operator basis of OPE,
as $u_*=d/2\in \{1/2,1,3/2,2,\cdots\}$.
The ambiguity induced by the $u=u_*$ renormalon in the momentum space is given by
\bea
\delta\tilde{C}_0(\tau)\big|_{u=u_*}
&=&\int d^3\vec{x}\,e^{-i\vec{\tau}\cdot\vec{x}}r^{2au'}\delta C_{0}(r^{-a})\big|_{u=u_*}\non
&=&-A_{u_*}\frac{4\pi}{\tau^{2au'+3}}\sin\big(\pi a(u_*+u')\big)\Gamma\big(2a(u_*+u')+2\big)\bigg(\frac{\LQ^2}{\tau^{2a}}\bigg)^{u_*}\,.
\label{deltatildeX}
\eea
It can be seen that 
$\delta\tilde{C}_0=0$ at $u_*=-u'+n/a$ ($n=0,1,2,...$) 
due to the sine factor in eq.~\eqref{deltatildeX}.
Owing to this property, 
we can eliminate the IR renormalons by a proper adjustment of the parameters $a$ and $u'$
due to the assumption (iii)\footnote{
Another source of singularities in the Borel plane is 
a proliferation of Feynman diagrams,
which causes the singularity at $u=4\pi n/b_0$ ($n$: integer).
If these singularities and IR renormalons are on the Borel plane,
the FTRS method cannot eliminate all the poles by any adjustment of $(a,u')$.
In the analyses of Sec.~\ref{sec3}, 
we do not consider the proliferation since 
it is much more distant from the origin on the Borel plane than IR renormalons,
and thus its impact on the perturbative calculation is considerably suppressed 
than the effect of IR renormalons.
}.

At the same time, however,
$\delta\tilde{C}_0$ contains the simple poles at $u_*=-u'-\frac{2\ell+3}{2a}$ ($\ell=0,1,2,...$).
These singularities give a sign-alternating divergent behavior to 
the perturbative expansion of eq.~\eqref{tildeX}.
We call these poles the ``artificial UV renormalons'' to distinguish them from 
the UV renormalon behavior originally contained in $C_0(Q)$.
Since UV renormalons in the Borel plane do not ruin the Borel integral [eq.~\eqref{Borel-def}],
the contribution from UV renormalons should not affect the calculation of $\tilde{C}_0$.
The full knowledge of the residue of these poles enables us to 
resum the contribution from the artificial UV renormalons in $\tilde{C}_0$.
The formulas necessary for the resummation are explained in the next section.

After the resummation of the artificial renormalons, 
$\tilde{C}_0(\tau)$ [eq.~\eqref{tildeX}] can be made as a renormalon-free series in the momentum space.
Then the accuracy of eq.~\eqref{FTRSdef} evaluated using perturbation theory 
can be systematically improved.

\subsubsection*{Perturbative construction of $\tilde{C}_0(\tau)$}
In this section, the procedure to construct $\tilde{C}_0(\tau)$ is given.
The perturbative expansion of $\tilde{C}_0(\tau)$ is determined by
\be
\big[\tilde{C}_0(\tau)\big]_{\rm PT}\equiv
\frac{4\pi}{\tau^{2au'+3}}\sum_{n=0}^\infty\tilde{c}_n(0)\alpha_s(\tau^a)^{n+1}
=\frac{4\pi}{\tau^{2au'+3}}F(\hat{H})\sum_{n=0}^\infty{c}_n(0)\alpha_s(\tau^a)^{n+1}\,,
\label{tildeX-PT}
\ee
where we set to $\mu=\tau^a$ and $\hat{H}=b_0\alfs(\tau^a)^2\big(\partial/\partial\alpha_s(\tau^a)\big)$.
$\tilde{c}_n(0)$ is the coefficient of the momentum-space series.
The $L_\tau=\log\big(\mu^2/\tau^{2a}\big)$ dependent coefficient $\tilde{c}_n(L_\tau)$
is obtained in the same manner as in eq.~\eqref{RGinv}.
The function $F(u)$ is given by
\be
F(u)=-\sin\big(\pi a(u+u')\big)\Gamma\big(2a(u+u')+2\big)\,,
\label{F(u)}
\ee
which is derived from eqs.~\eqref{RGinv} and \eqref{tildeX}.
As mentioned in the last section, 
we must resum the artificial UV renormalons in $F(u)$ 
to eliminate the contribution from them to $\tilde{c}_n$.
Procedure to resum the artificial UV renormalons is as follows:
we separate $F(u)$ into two parts,
\be
F(u)=F^{\rm subt}(u)+F^{\rm UV}(u)\,,
\ee
where $F^{\rm UV}(u)$ is the sum of all the poles at $u_*=-u'-\frac{2\ell+3}{2a}$ ($\ell=0,1,2,...$).
Then $F^{\rm UV}(u)$ is given explicitly by 
\be
F^{\rm UV}(u)=\sin(\pi a(u+u'))\sum_{\ell=0}^\infty\frac{1}{(2\ell+1)!\big(2a(u+u')+2\ell+3\big)}.
\ee
$F^{\rm subt}(u)$ is given by 
\be
F^{\rm subt}(u)=F(u)-F^{\rm UV}(u)\,,
\ee
where all the artificial UV renormalons are subtracted.
First, we construct the momentum-space series,
from which all the UV renormalons are subtracted, as
\be
\big[\tilde{C}_0(\tau)\big]_{\rm subt}
\equiv\frac{4\pi}{\tau^{2au'+3}}\sum_{n=0}^\infty\tilde{c}^{\rm subt}_n\alpha_s(\tau^a)^{n+1}
=\frac{4\pi}{\tau^{2au'+3}}F^{\rm subt}(\hat{H})\sum_{n=0}^\infty{c}_n\alpha_s(\tau^a)^{n+1}\,.
\label{tildeX-subt}
\ee
On the right hand side of eq.~\eqref{tildeX-subt},
the operation of $F^{\rm subt}(\hat{H})$ is calculated as follows:
\bea
&&F^{\rm subt}(\hat{H})\sum_{n=0}^\infty{c}_n(0)\alpha_s(\tau^a)^{n+1}
=\sum_{m=0}^\infty F_m\hat{H}^m\sum_{n=0}^\infty{c}_n(0)\alpha_s(\tau^a)^{n+1}\non
&&~~~~~~
=\sum_{n,m=0}^\infty F_mb_0^m{c}_n(0)\frac{(n+m)!}{n!}\alpha_s(\tau^a)^{n+m+1}\non
&&~~~~~~
=F_0\big(c_0(0)\alpha_s(\tau^a)+c_1(0)\alpha_s(\tau^a)^2+c_2(0)\alpha_s(\tau^a)^3+\cdots\big)\non
&&~~~~~~
+F_1b_0\big(c_0(0)\alpha_s(\tau^a)^2+2c_1(0)\alpha_s(\tau^a)^3+3c_2(0)\alpha_s(\tau^a)^4+\cdots\big)\non
&&~~~~~~
+F_2b_0^2\big(2c_0(0)\alpha_s(\tau^a)^2+6c_1(0)\alpha_s(\tau^a)^3+6c_2(0)\alpha_s(\tau^a)^4+\cdots\big)\non
&&~~~~~~~~
+\cdots\,,
\label{FH-expl}
\eea
where $F_m=\frac{1}{m!}\frac{\partial^m}{\partial u^m}F^{\rm subt}(u)\Big|_{u=0}$.
The first few coefficients $\tilde{c}^{\rm subt}_n(0)$ 
are given explicitly by
\bea
&&\tilde{c}^{\rm subt}_0(0)=F_0c_0(0)\,,\\
&&\tilde{c}^{\rm subt}_1(0)=F_0c_1(0)+F_1c_0(0)b_0\,,\\
&&\tilde{c}^{\rm subt}_2(0)=F_0c_2(0)+2F_12c_1(0)b_0+2F_2c_0(0)b_0^2\,.
\label{FH-expl2}
\eea
For an integer $k$, 
$\tilde{c}^{\rm subt}_k(0)$ is written by the combination of $c_n(0)$ $(n=0,\,1,\,\cdots,\,k)$ and 
$b_0$.
In the practical situation, 
the perturbative expansion of $C_0(Q)$ is known up to order $\alfs^{k+1}$.
Then we construct $\big[\tilde{C}_0(\tau)\big]_{\rm subt}$ from 
the first $(k+1)$ terms of the perturbative expansion of $C_0(Q)$,
given by
\be
\big[\tilde{C}_0(\tau)\big]_{\rm subt}^{(k)}=\frac{4\pi}{\tau^{2au'+3}}\sum_{n=0}^k\tilde{c}^{\rm subt}_n(0)\alpha_s(\tau^a)^{n+1}.
\label{tildeXsubt-k}
\ee
The important feature of eq.~\eqref{tildeXsubt-k} is that
$\tilde{c}_n(0)$ does not have the 
factorial divergent behavior due to the proper choice of the parameters of Fourier transform and 
subtraction of the artificial UV renormalons, i.e.,
eq.~\eqref{tildeXsubt-k} is free of all the renormalons.

Secondly, we resum the UV renormalons in the counterpart as follows,\footnote{
Compared to ref.~\cite{Hayashi:2021vdq}, the treatment of the artificial UV renormalons is modified 
to resum all the UV renormalons
while keeping the suppression of IR renormalons by the sine factor.}
\bea
&&\big[\tilde{C}_0(\tau)\big]_{\rm resum}
=\frac{4\pi}{\tau^{2au'+3}}F^{\rm UV}(\hat{H})\sum_{n=0}^\infty{c}_n\alpha_s(\tau^a)^{n+1}\non
&&=\frac{4\pi}{\tau^{2au'+3}}\sin(\pi a(\hat{H}+u'))
\sum_{\ell=0}^\infty\frac{1}{(2\ell+1)!\big(2a(\hat{H}+u')+2\ell+3\big)}
\sum_{n=0}^\infty{c}_n\alpha_s(\tau^a)^{n+1}\non
&&=\frac{4\pi}{\tau^{2au'+3}}\sin(\pi a(\hat{H}+u'))
\sum_{\ell=0}^\infty\frac{1}{(2\ell+1)!}\int_0^1dv\,v^{2a(\hat{H}+u')+2\ell+2}
\sum_{n=0}^\infty{c}_n\alpha_s(\tau^a)^{n+1}\non
&&=\frac{4\pi}{\tau^{2au'+3}}\int_0^1dv\,\sinh(v)v^{2au'+1}
\sum_{n=0}^\infty\tilde{c}_n^{\rm resum}\alpha_s\big((\tau/v)^a\big)^{n+1}.
\eea
In the last equality, 
we exchange the order of the integration and summation.
The relation among $\tilde{c}^{\rm resum}_n(0)$ and $c_n(0)$ is given by
\be
\sum_{n=0}^\infty\tilde{c}^{\rm resum}_n\alpha_s^{n+1}
=\sin(\pi a(\hat{H}+u'))\sum_{n=0}^\infty{c}_n\alpha_s^{n+1}\,.
\ee
Each coefficient $\tilde{c}^{\rm resum}_n$ can be obtained 
in the same manner as in eq.~\eqref{FH-expl}.
We note that using RGE, the contribution from the UV renormalons is absorbed into 
the running of $\alfs$.\footnote{
Since it is the same pole singularity that gives alternating divergent behavior beyond the large-$\beta_0$ approximation, the formula to resum the artificial UV renormalon holds beyond the large-$\beta_0$ approximation.
}
From the first $(k+1)$ terms of the perturbative expansion of $C_0(Q)$,
we construct $\tilde{C}_0\,_{\rm resum}$ up to $\alfs^{k+1}$,
given by
\be
\tilde{C}_0(\tau)_{\rm resum}^{(k)}
=\frac{4\pi}{\tau^{2au'+3}}\int_0^1dv\sinh(v)v^{2au'+1}
\sum_{n=0}^k\tilde{c}^{\rm resum}_n\alpha_s((\tau/v)^a)^{n+1}.
\label{tildeXresum-k}
\ee

In short, after the resummation of the artificial UV renormalons, 
the momentum-space series is given by 
\bea
&&\tilde{C}_0(\tau)^{(k)}
=\tilde{C}_0(\tau)_{\rm subt}^{(k)}+\tilde{C}_0(\tau)_{\rm resum}^{(k)}\non
&&=\frac{4\pi}{\tau^{2au'+3}}\sum_{n=0}^k\Bigg[\tilde{c}^{\rm subt}_n\alpha_s(\tau^a)^{n+1}
+\int_0^1dv\sinh(v)v^{2au'+1}\,\tilde{c}^{\rm resum}_n\alpha_s((\tau/v)^a)^{n+1}\Bigg].
\label{tildeX-k}
\eea

\subsubsection*{Perturbative computation of $\big[C_0(Q)\big]_{\rm FTRS}$}
\begin{figure}[t]
\centering
\includegraphics[width=5cm]{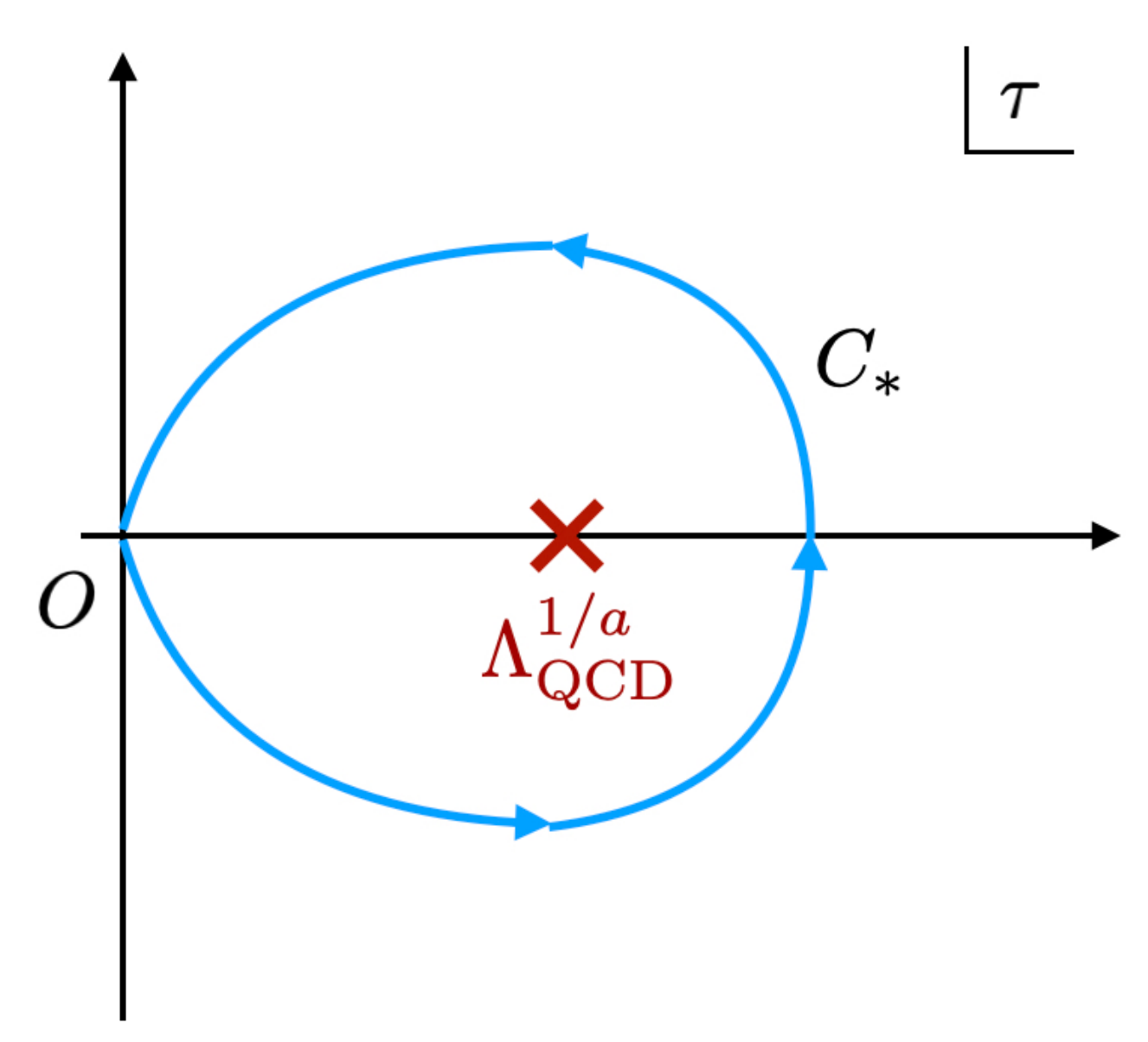}
\caption{Integration contour $C_*$
surrounding the Landau singularity 
of the one-loop running coupling constant.
[$\alfs(q)$ diverges at $q=\LQ$.]}
\label{fig:contour}
\end{figure}

In this section, using eq.~\eqref{tildeX-k},
we compute the principal value of the momentum integral (PV integral) in eq.~\eqref{FTRSdef}
with the momentum-space series truncated at ${\cal O}(\alfs^{k+1})$.
Here, we separate again $\tilde{C}_0(\tau)^{(k)}$ into two parts,
\be
\big[C_0(Q)\big]^{(k)}_{\rm FTRS}=\big[C_0(Q)\big]^{(k)}_{\rm subt}+\big[C_0(Q)\big]^{(k)}_{\rm resum}\,.
\ee
The first term $\big[C_0(Q)\big]^{(k)}_{\rm subt}$ denotes the UV renormalon-subtracted part given by
\be
\big[C_0(Q)\big]^{(k)}_{\rm subt}=\frac{r^{-2au'-1}}{2\pi^2}\int_{0,\rm PV}^\infty \!\!d\tau\,\tau\sin\big(\tau r\big)\big[\tilde{C}_0(\tau)\big]_{\rm subt}^{(k)}\,,
\ee
and the resummation part $\big[C_0(Q)\big]^{(k)}_{\rm resum}$ is given by
\be
\big[C_0(Q)\big]^{(k)}_{\rm resum}=\frac{r^{-2au'-1}}{2\pi^2}\int_{0,\rm PV}^\infty \!\!d\tau\,\tau\sin\big(\tau r\big)\big[\tilde{C}_0(\tau)\big]_{\rm resum}^{(k)}\,.
\label{XFTRS-resum}
\ee
The PV integral can be reduced to the following form
by deforming the integration contour [see eqs.(28) -- (34) in ref.~\cite{Hayashi:2020ylq}],
\bea
&&\big[C_0(Q)\big]^{(k)}_{\rm FTRS}\non
&&
=\frac{-r^{-2au'-1}}{2\pi^2}\int_{0}^\infty \!\!dt\,t\exp(-t r){\rm Im}\Big[\big[\tilde{C}_0(it)\big]_{\rm subt}^{(k)}\Big]+\frac{r^{-2au'-1}}{4\pi^2i}\int_{C_*}\!\!d\tau\,\tau\cos(\tau r)\big[\tilde{C}_0(\tau)\big]_{\rm subt}^{(k)}\non
&&
+\frac{-r^{-2au'-1}}{2\pi^2}\int_{0}^\infty \!\!dt\,t\exp(-t r){\rm Im}\Big[\big[\tilde{C}_0(it)\big]_{\rm resum}^{(k)}\Big]+\frac{r^{-2au'-1}}{4\pi^2i}\int_{C_*}\!\!d\tau\,\tau\cos(\tau r)\big[\tilde{C}_0(\tau)\big]_{\rm resum}^{(k)}\,,\non
\label{X-FTRS-k}
\eea
where $C_*$ denotes the integration contour surrounding the pole singularity
induced by $\alfs(\tau^a)$ (see Fig.~\ref{fig:contour}).

Since $\big[\tilde{C}_0(\tau)\big]_{\rm resum}^{(k)}$ includes the $v$-integral, 
we need to evaluate the two-parameter integral.
The third term of eq.~\eqref{X-FTRS-k} is given by
\bea
&&\frac{-r^{-2au'-1}}{2\pi^2}\int_{0}^\infty \!\!dt\,t\exp(-t r){\rm Im}\Big[\big[\tilde{C}_0(it)\big]_{\rm resum}^{(k)}\Big]\non
&&=\frac{-2r^{-2au'-1}}{\pi}\int_0^1dv\,\sinh(v)v^{2au'+1}
\int_{0}^\infty \!\!dt\,\frac{\exp(-t r)}{t^{2au'+2}}\sum_{n=0}^k\tilde{c}_n^{\rm resum}{\rm Im}\big[(-i)^{2au'+1}\alpha_s\big((it/v)^a\big)^{n+1}\big]\non
&&=\frac{-2r^{-2au'-1}}{\pi}\int_0^1dv\,\sinh(v)
\int_{0}^\infty \!\!dt\,\frac{\exp(-t v r)}{t^{2au'+2}}\sum_{n=0}^k\tilde{c}_n^{\rm resum}{\rm Im}\big[(-i)^{2au'+1}\alpha_s((it)^a)^{n+1}\big]\,.
\label{vint1}
\eea
In the last equality, we changed the integration variable as $t\to tv$.
The last term of eq.~\eqref{X-FTRS-k} is given by
\bea
&&\frac{r^{-2au'-1}}{4\pi^2i}\int_{C_*}\!\!d\tau\,\tau\cos(t r)\big[\tilde{C}_0(\tau)\big]_{\rm resum}^{(k)}\non
&&=\frac{r^{-2au'-1}}{\pi i}\int_0^1dv\,\sinh(v)v^{2au'+1}
\int_{C_*} \!\!d\tau\,\frac{\cos(\tau r)}{\tau^{2au'+2}}\sum_{n=0}^k\tilde{c}_n^{\rm resum}\alpha_s\big((\tau/v)^a\big)^{n+1}\non
&&=\frac{r^{-2au'-1}}{\pi i}\int_0^1dv\,\sinh(v)
\int_{C_*} \!\!d\tau\,\frac{\cos(\tau v r)}{\tau^{2au'+2}}\sum_{n=0}^k\tilde{c}_n^{\rm resum}\alpha_s\big(\tau^a\big)^{n+1}\,.
\label{vint2}
\eea
In the last equality, we changed the integration variable as $\tau\to \tau v$.

We can evaluate the $v$-integral in eqs.~\eqref{vint1} and \eqref{vint2} analytically.
To avoid cumbersome expressions,
the result of the $v$-integral is not presented here,
but we show the explicit forms in the analyses in the next section.

In the same way, we can define the ambiguous part $\delta C_0(Q)$ subtracted by taking 
the principal value of the momentum integral.
It is given by
\be
\big[\delta C_0(Q)\big]^{(k)}_{\rm FTRS}
=\frac{r^{-2au'-1}}{4\pi^2i}\int_{C_*}\!\!d\tau\,\tau\sin(\tau r)\Big[\big[\tilde{C}_0(\tau)\big]_{\rm subt}^{(k)}+\big[\tilde{C}_0(\tau)\big]_{\rm resum}^{(k)}\Big].
\ee
The $v$-integral included in the $\big[\tilde{C}_0(\tau)\big]_{\rm resum}^{(k)}$ can be evaluated 
by the same procedure as in eq.~\eqref{vint2}.

\section{Analyses of simplified models}
\label{sec3}

In this sections, we perform analyses of simplified toy models. 
We use the one-loop beta function ($b_i=0$ for $i=1,2,3,\cdots$)
to show the explicit process of the analytical computation.
Then the running coupling constant is given by
\be
\alfs(Q)=\frac{1}{b_0\log\big(Q^2/\LQ^2\big)},
\label{as-exact}
\ee
where $\LQ$ is the Landau pole of the QCD coupling.
\subsection{Model 1: $\alpha_s(Q)$}
\label{sec3-1}
In this section, the toy model is the running coupling itself\footnote{
The difference from the analysis in Sec.~2.3 of ref.~\cite{Hayashi:2021vdq}
is the formula for the resummation of the artificial UV renormalons.
}.
The perturbative expansion of $\alfs(Q)$ is given by
\be
\big[\alfs(Q)\big]_{\rm PT}=\sum_{n=0}^\infty b_0^n\log^n(\mu^2/Q^2)\alfs^{n+1},
\label{as-PT}
\ee 
which is a convergent series if $|b_0\alpha_s\log(\mu^2/Q^2)|<1$.
In this analysis, we investigate 
what happens if the FTRS method is applied to the convergent series.
A naive expectation is that 
it does not cause anything because there is no singularity in the Borel plane 
for the convergent series.
We compare the result of our method with the result of the resummed form [eq.~\eqref{as-exact}]
to check such an expectation.

Here we adjust the parameters to $a=1$ and $u'=-1$.
Then the Fourier transform of $\alfs(Q)$ is given by
\be
\tilde{\alfs}(\tau)=\frac{4\pi}{\tau}G(\hat{H})\alfs(\tau)\,,
\quad
G(u)=\sin(\pi u)\Gamma(2u)\,, 
\label{tildeas}
\ee
where $\hat{H}=b_0\alfs(\tau)^2\big(\partial/\partial \alfs(\tau)\big)$.
It can be seen that $G(u)$ contains the UV renormalon poles at $u=-\frac{\ell+1}{2}$ 
for $\ell=0,1,2,\cdots$.

According to the procedure in Sec.\ref{sec2}, 
we separate the momentum-space series into the part without artificial UV renormalons
and the part resumming the contribution from the UV renormalons.
We define the function $G^{\rm subt}(u)$ by
\be
G^{\rm subt}(u)=\sin(\pi u)\Bigg[\Gamma\big(2u\big)+\sum_{\ell=0}^\infty\frac{1}{(2\ell+1)!(2u+2\ell+1)}\Bigg]\,.
\ee
Then we obtain the momentum-spcae series without the artificial UV renormalons,
\bea
&&\big[\tilde{\alfs}(\tau)\big]_{\rm subt}
=\frac{4\pi}{\tau}G^{\rm subt}(\hat{H})\alfs(\tau)
=\sum_{n=0}^\infty\tilde{a}_n^{\rm subt}\alpha_s\big(\tau\big)^{n+1}\non
&&\approx\frac{4\pi}{b_0\tau}\bigg[1.57 x_\tau + 1.51 x_\tau^2 - 5.54 x_\tau^3 - 7.44 x_\tau^4 + 25.7 x_\tau^5 + 14.7 x_\tau^6-232x_\tau^7+\cdots\bigg]\,,
\label{tildeas-subt}
\eea
where $x_\tau=b_0\alfs(\tau)$.
It can be seen that it does not show a factorial divergence due to renormalons, 
but it appears not to be convergent.
Below, we will see that 
the convergence of the result of the FTRS method is recovered by taking the principal value of 
the inverse Fourier integral.


On the other hand, 
the resummation part is given by
\bea
\big[\tilde{\alfs}(\tau)\big]_{\rm resum}
&=&-\frac{4\pi}{\tau}\int_0^1\frac{dv}{v}\,\sinh(v)\sin(\pi\hat{H})
\alpha_s\big(\tau/v\big)\non
&=&
\frac{4\pi}{\tau}\int_0^1\frac{dv}{v}\,\sinh(v)\sum_{n=0}^\infty\tilde{a}_n^{\rm resum}\alpha_s\big(\tau/v\big)^{n+1},
\label{tildeas-resum}
\eea
where $\tilde{a}_n^{\rm resum}$ is determined from the following relation,
\bea
&&\sum_{n=0}^\infty\tilde{a}^{\rm resum}_n\alpha_s^{n+1}
=-\sin(\pi\hat{H})\alpha_s
=\sum_{m=0}^\infty\frac{(-1)^{m+1}}{(2m+1)!}(\pi \hat{H})^{2m+1}\alfs\non
&&~~
=\sum_{m=0}^\infty{(-1)^{m+1}}(\pi b_0)^{2m+1}\alfs^{2m+2}=\frac{-\pi b_0\alfs^2}{1+\pi^2b_0^2\alfs^2}\,,
\eea
and given by
\be
\tilde{a}^{\rm resum}_{2m}=0\quad{\rm for}\quad m=0,1,2,\cdots\,,
\ee
\be
\tilde{a}^{\rm resum}_{2m+1}=(-1)^{m+1}(\pi b_0)^{2m+1}\quad{\rm for}\quad m=0,1,2,\cdots\,.
\ee
It can be seen that $\tilde{a}_n^{\rm resum}$ does not have a factorial divergent behavior
owing to the resummation using RGE.

In order to consider the situation similar to the practical application (cf. ref.~\cite{Hayashi:2020ylq,Hayashi:2021vdq}), 
we truncate the momentum-space series at ${\cal O}(\alfs^{k+1})$.
(This manipulation corresponds to the N$^k$LL analysis.)
Then we obtain the one-parameter integral form of $\big[\alfs(Q)\big]_{\rm FTRS}$
in the case of the truncation at ${\cal O}(\alfs^{k+1})$.
It is given by
\be
\big[\alfs(Q)\big]^{(k)}_{\rm FTRS}
=\big[\alfs(Q)\big]^{(k)}_{\rm subt}+\big[\alfs(Q)\big]^{(k)}_{\rm resum}\,,
\ee
where each part is given by
\bea
&&\big[\alfs(Q)\big]^{(k)}_{\rm subt}
=\frac{2}{\pi Q}\int_{0}^\infty \!\!dt\,\exp(-t/Q)\sum_{n=0}^k \tilde{a}_n^{\rm subt}
\,{\rm Re}\big[\alfs(it)^{k+1}\big]\non
&&~~~~~~~~~~~~~~~~~
+\frac{1}{\pi Qi}\int_{C_*}\!\!d\tau\,\cos(\tau/Q)\sum_{n=0}^k \tilde{a}_n^{\rm subt}
\alfs(\tau)^{n+1}\,,
\label{as-subtk}
\eea
and
\bea
&&\big[\alfs(Q)\big]^{(k)}_{\rm resum}
=\frac{2}{\pi Q}\int_0^1dv\,\sinh(v)\int_{0}^\infty \!\!dt\,\exp(-tv/Q)\sum_{n=0}^k 
\tilde{a}_n^{\rm resum}
\,{\rm Re}\big[\alfs(it)^{n+1}\big]\non
&&~~~~~~~~~~~~~~~~~
+\frac{1}{\pi Qi}\int_0^1dv\,\sinh(v)\int_{C_*}\!\!d\tau\,\cos(\tau v/Q)\sum_{n=0}^k \tilde{a}_n^{\rm resum}\alfs(\tau)^{n+1}\non
&&=\frac{2}{\pi Q}\int_{0}^\infty \!\!dt\,\frac{e^{-t/Q}\big(\cosh(1)+t/Q\sinh(1)\big)-1}{1-t^2/Q^2}\sum_{n=0}^k \tilde{a}_n^{\rm resum}
\,{\rm Re}\big[\alfs(it)^{n+1}\big]\non
&&
+\frac{1}{\pi Qi}\int_{C_*}\!\!d\tau\,\frac{-1+\cos(\tau/Q)\cosh(1)+\tau/Q\sin(\tau/Q)\sinh(1)}{1+\tau^2/Q^2}\sum_{n=0}^k \tilde{a}_n^{\rm resum}\alfs(\tau)^{n+1}\,.\non
\label{as-resumk}
\eea

We discuss the convergence of eqs.~\eqref{as-subtk} and \eqref{as-resumk}.
Both of them consist of the exponential-damping integral and closed-path integral.
The $n$-th term of exponential-damping integral is proportional to
\be
I_n=\frac{1}{Q}\int_0^\infty dt\,\exp(-t/Q)\,{\rm Re}\Bigg[\frac{1}{\big(\log(t^2/\LQ^2)+i\pi\big)^n}\Bigg]\,.
\ee
We can set the upper limit for the absolute value of $I_n$ as
$|I_n|\leq1/\pi^n$
which considerably suppresses each term of the expansion.\footnote{
It is confirmed that $I_n$ for large $n$ is more suppressed than the power-like upper limit
by the numerical computation.
Detailed discussion is given in App.G of ref.~\cite{Hayashi:2021vdq}.
}
For the $n$-th term of the closed-path integral part, 
the suppression factor of order $1/n!$ stems from the residue of $\alpha_s(\tau)^{n+1}$
at $\tau=\LQ$.
Hence, the result of the FTRS method shows a convergent behavior
as the truncation order $k$ increases.\footnote{
It is impossible to discuss the convergence of this method parametrically since 
the integral $I_n$ cannot be evaluated analytically,
and $\tilde{a}_n^{\rm subt}$ ($\tilde{a}_n^{\rm resum}$), the each coefficient of $I_n$ in eq.~\eqref{as-subtk} (eq.~\eqref{as-resumk}) is complicated due to the separation of generated UV renormalon from $\tilde{a}_n$.
}

\begin{figure}[t]
\begin{center}
\includegraphics[width=14cm]{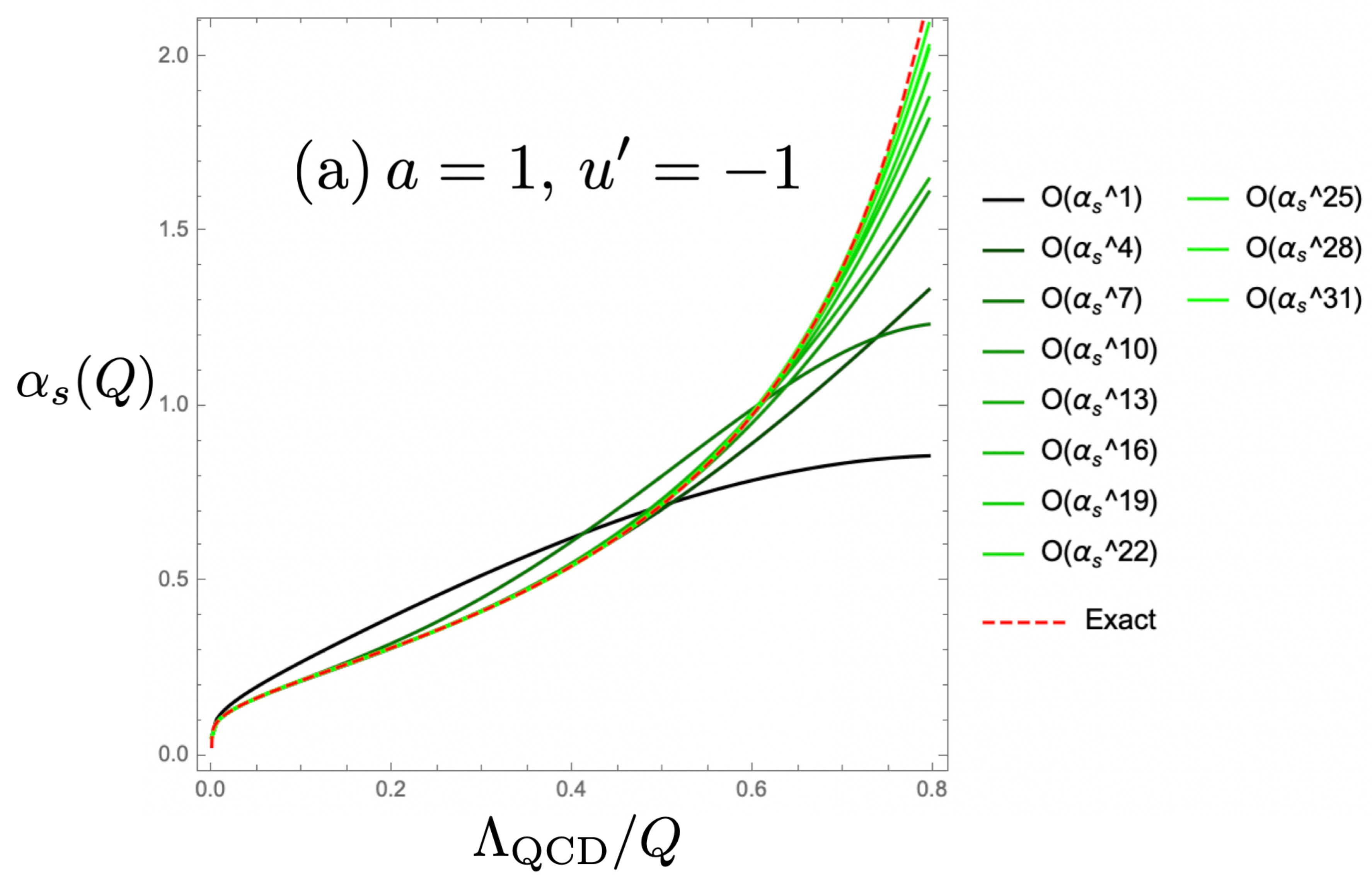}
\end{center}
\vspace*{-5mm}
\caption{\small Comparison of the exact result (dashed red line) 
and the FTRS method for $\alpha_s(Q)$ (green lines with gradation)
with the parameters adjusted to $(a,u')=(1,-1)$.
We set $b_0=1$.
In the momentum space, we truncate the series at ${\cal O}(\alfs^{k+1})$
for $k=1,\cdots,31$.
}
\label{Fig:as-analysis-a}
\vspace*{-4mm}
\end{figure}

We compare the result of the FTRS method with
the exact value of $\alfs(Q)$ [eq.~\eqref{as-exact}].
Fig.~\ref{Fig:as-analysis-a} shows the result. 
The dashed red line represents the exact value.
The green lines with gradation represent the result of the FTRS method
truncated at ${\cal O}(\alfs^{k+1})$ for $k=1,4,\cdots,31$.
As higher order terms are incorporated in the momentum series,
the result of the FTRS method approaches the exact result gradually.
This result shows that our method does not affect the convergence
for originally convergent series.
We note that the result of the FTRS method is not the same as 
the truncated result with fixed $\mu$ of the original series [eq.~\eqref{as-PT}].

\begin{figure}[t]
\begin{center}
\includegraphics[width=14cm]{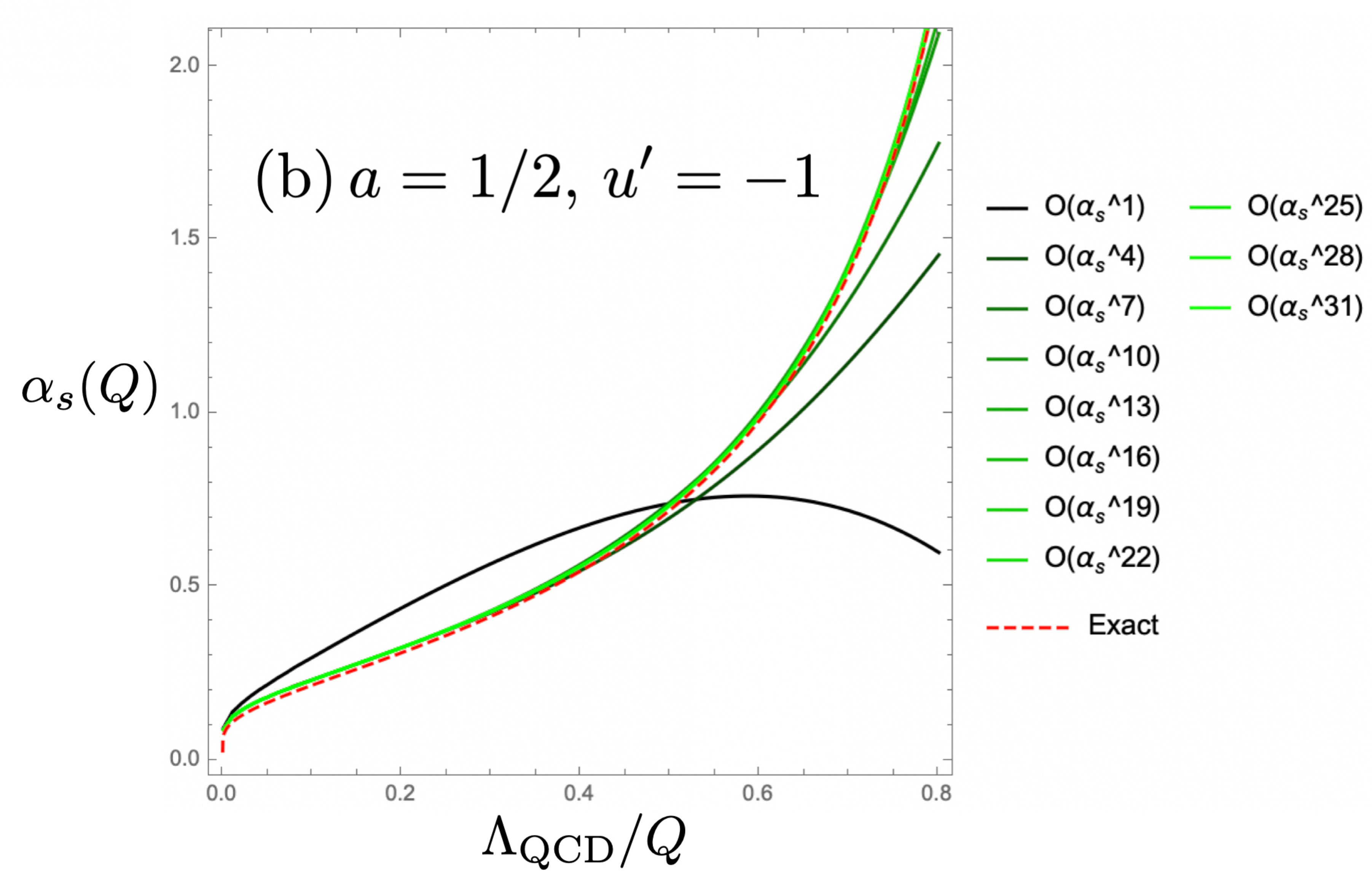}
\end{center}
\vspace*{-5mm}
\caption{\small Comparison of the exact result (dashed red line) 
and the FTRS method for $\alpha_s(Q)$ (green lines with gradation)
with the parameters adjusted to $(a,u')=(1/2,-1)$.
We set $b_0=1$.
In the momentum space, we truncate the series at ${\cal O}(\alfs^{k+1})$
for $k=1,\cdots,31$.
}
\label{Fig:as-analysis-b}
\vspace*{-4mm}
\end{figure}

In addition, we investigate the situation with the other choice of the parameters $(a,u')=(1/2,-1)$.
Since $\alpha_s(Q)$ is free of IR renormalons, 
any choice of $(a,u')$ should not ruin the convergence of the result of
the FTRS method.
As in Fig.~\ref{Fig:as-analysis-b}, 
we confirm the convergence of the result of the FTRS method in this case.
We can see more rapid convergence for the case with $(a,u')=(1/2,-1)$ than with $(a,u')=(1,-1)$.
This is because the $\LQ/Q$ expansion of the closed-path integral part of the former
takes the form of $\sum_{m=0}^\infty A_m(\LQ/Q)^{4m}$,
which is more convergent series than that of the latter which takes the form of 
$\sum_{m=0}^\infty A'_m(\LQ/Q)^{2m}$.

\subsection{Model 2: $u=1,\,2$ renormalons}
\label{sec3-2}
In this section, we examine another model $X_1(Q)$ defined by the following Borel transform 
\be
B_{X_1}(u)=\frac{e^{L_Qu}}{(1-u)(2-u)}
=\sum_{n=0}^\infty \frac{d_n(L_Q)}{n!}\bigg(\frac{u}{b_0}\bigg)^n\,.
\ee
This function contains the pole singularities at $u=1$ and $u=2$ in the complex Borel plane.
Thus, the perturbative series of $X_1(Q)$ is affected by the renormalons, 
and it is given by
\bea
&&\big[X_1(Q)\big]_{\rm PT}=\sum_{n=0}^\infty d_n(L_Q)\alfs^{n+1}
=\sum_{n=0}^\infty \frac{2^{n+1}-1}{2^{n+1}}n!b_0^n\alfs(Q)^{n+1}\non
&&\approx\frac{1}{b_0}\bigg[0.5x + 0.75x^2 + 
 1.75x^3 + 5.63x^4 + 
 23.3x^5 + 118x^6 + 
 714x^7+\cdots\bigg]\,,
\eea
where $x=b_0\alpha_s(Q)$.
Using the PV prescription, eq.~\eqref{X-PV}, we can calculate the finite exact value from $B_{X_1}(u)$,
given by
\bea
\big[X_1(Q)\big]_{\rm PV}
&=&\frac{e^{-\frac{1}{b_0\alfs(Q)}}}{b_0}{\rm Ei}\Big(\frac{1}{b_0\alfs(Q)}\Big)-\frac{e^{-\frac{2}{b_0\alfs(Q)}}}{b_0}{\rm Ei}\Big(\frac{2}{b_0\alfs(Q)}\Big),
\label{X1-PV}
\eea
where ${\rm Ei}$ stands for the exponential integral given by
${\rm Ei}(z)=-\int_{-z}^\infty\frac{dt}{t}e^{-t}$.
We study how the renormalon subtraction for $X_1(Q)$ works using the FTRS method.
The result of the method is compared with the PV-prescription result [eq.~\eqref{X1-PV}].

In order to eliminate the $u=1$ and $u=2$ renormalons, 
we choose the parameters as $a=1$ and $u'=-1$.
Then the Fourier transform of $X_1(Q)$ is given by
\be
\tilde{X}_1(\tau)=\frac{4\pi}{\tau}G(\hat{H})\sum_{n=0}^\infty d_n\alfs(\tau)^{n+1},
\label{tildeX1}
\ee
where $G(u)$ is given by eq.~\eqref{tildeas}.
Due to the sine factor in $G(u)$,
eq.~\eqref{tildeX1} does not contain IR renormalons.

On the other hand, the UV renormalons in the momentum space series
genetated from the singularity of $G(u)$ should be separated and resummed.
The formula is given by
\be
\tilde{X}_1(\tau)=\big[\tilde{X}_1(\tau)\big]_{\rm subt}+\big[\tilde{X}_1(\tau)\big]_{\rm resum}.
\label{tildeX1-sep}
\ee
The first term in eq.~\eqref{tildeX1-sep} is given by
\bea
&&\big[\tilde{X}_1(\tau)\big]_{\rm subt}
=\frac{4\pi}{\tau}G^{\rm subt}(\hat{H})\sum_{n=0}^\infty d_n\alfs(\tau)^{n+1}
=\frac{4\pi}{\tau}\sum_{n=0}^\infty \tilde{d}_n^{\rm subt}\alfs(\tau)^{n+1}\non
&\approx&\frac{4\pi}{b_0\tau}\bigg[0.785x_\tau + 1.93x_\tau^2 + 
 2.24x_\tau^3 + 0.564x_\tau^4 + 
 2.81x_\tau^5 + 22.8x_\tau^6 + 
 46.6x_\tau^7+\cdots\bigg]\,,
\eea
where $x_\tau=b_0\alpha_s(\tau)$.
It can be seen that 
$\tilde{d}_n^{\rm subt}$ is significantly suppressed compared to
the factorial behavior of ${d}_n$,
due to the elimination of the IR renormalons and separation of the generated UV renormalons
in the momentum space.

The second term in eq.~\eqref{tildeX1-sep} is given by
\bea
\big[\tilde{X}_1(\tau)\big]_{\rm resum}
&=&\frac{4\pi}{\tau}\int_0^1\frac{dv}{v}\,\sinh(v)\sum_{n=0}^\infty\tilde{d}_n^{\rm resum}\alpha_s\big(\tau/v\big)^{n+1},
\label{tildeX1-resum}
\eea
where $\tilde{d}_n^{\rm resum}$ is determined from the following relation
\bea
&&\sum_{n=0}^\infty\tilde{d}_n^{\rm resum}\alpha_s^{n+1}
=-\sin(\pi\hat{H})\sum_{n=0}^\infty d_n\alpha_s^{n+1}
=\sum_{\ell=0}^\infty\frac{(-1)^{\ell+1}}{(2\ell+1)!}(\pi \hat{H})^{2\ell+1}\sum_{n=0}^\infty d_n\alpha_s^{n+1}\non
&&~~~~~~~~~~~~~~~~~~~
=\sum_{\ell=0}^\infty\sum_{n=0}^\infty (-1)^{\ell+1}
\frac{(n+2\ell+1)!}{(2\ell+1)!n!}(\pi b_0)^{2\ell+1}d_n\alpha_s^{n+2\ell+2}\non
&&\approx
\frac{1}{b_0}\Big[-1.57 x^2 - 4.71x^3 - 0.990 x^4 + 22.3 x^5 + 24.4 x^6 - 115 x^7+\cdots\Big]
 \,,
\eea
where $x=b_0\alfs$.
We can see that $\tilde{d}^{\rm resum}_n$ does not grow factorially,
because of the resummation of the UV renormalons.

Now, we show the one-parameter integral form of $\big[X_1(Q)\big]_{\rm FTRS}$
in the case of the truncation at ${\cal O}(\alfs^{k+1})$.
It is given by
\be
\big[X_1(Q)\big]_{\rm FTRS}^{(k)}
=\big[X_1(Q)\big]_{\rm subt}^{(k)}+\big[X_1(Q)\big]_{\rm resum}^{(k)},
\label{X1FTRS}
\ee
where each part is given by
\bea
&&\big[X_1(Q)\big]^{(k)}_{\rm subt}
=\frac{2}{\pi Q}\int_{0}^\infty \!\!dt\,\exp(-t/Q)\sum_{n=0}^k \tilde{d}_n^{\rm subt}
\,{\rm Re}\big[\alfs(it)^{n+1}\big]\non
&&~~~~~~~~~~~~~~~~~
+\frac{1}{\pi Qi}\int_{C_*}\!\!d\tau\,\cos(\tau/Q)\sum_{n=0}^k \tilde{d}_n^{\rm subt}
\alfs(\tau)^{n+1}\,,
\label{X1subt}
\eea
and
\bea
&&\big[X_1(Q)\big]^{(k)}_{\rm resum}
=\frac{2}{\pi Q}\int_0^1dv\,\sinh(v)\int_{0}^\infty \!\!dt\,\exp(-tv/Q)\sum_{n=0}^k 
\tilde{d}_n^{\rm resum}
\,{\rm Re}\big[\alfs(it)^{n+1}\big]\non
&&~~~~~~~~~~~~~~~~~
+\frac{1}{\pi Qi}\int_0^1dv\,\sinh(v)\int_{C_*}\!\!d\tau\,\cos(\tau v/Q)\sum_{n=0}^k \tilde{d}_n^{\rm resum}\alfs(\tau)^{n+1}\non
&&=\frac{2}{\pi Q}\int_{0}^\infty \!\!dt\,\frac{e^{-t/Q}\big(\cosh(1)+t/Q\sinh(1)\big)-1}{1-t^2/Q^2}\sum_{n=0}^k \tilde{d}_n^{\rm resum}
\,{\rm Re}\big[\alfs(it)^{n+1}\big]\non
&&
+\frac{1}{\pi Qi}\int_{C_*}\!\!d\tau\,\frac{-1+\cos(\tau/Q)\cosh(1)+\tau/Q\sin(\tau/Q)\sinh(1)}{1+\tau^2/Q^2}\sum_{n=0}^k \tilde{d}_n^{\rm resum}\alfs(\tau)^{n+1}\,.\non
\label{X1resum}
\eea

\begin{figure}[t]
\begin{center}
\includegraphics[width=14cm]{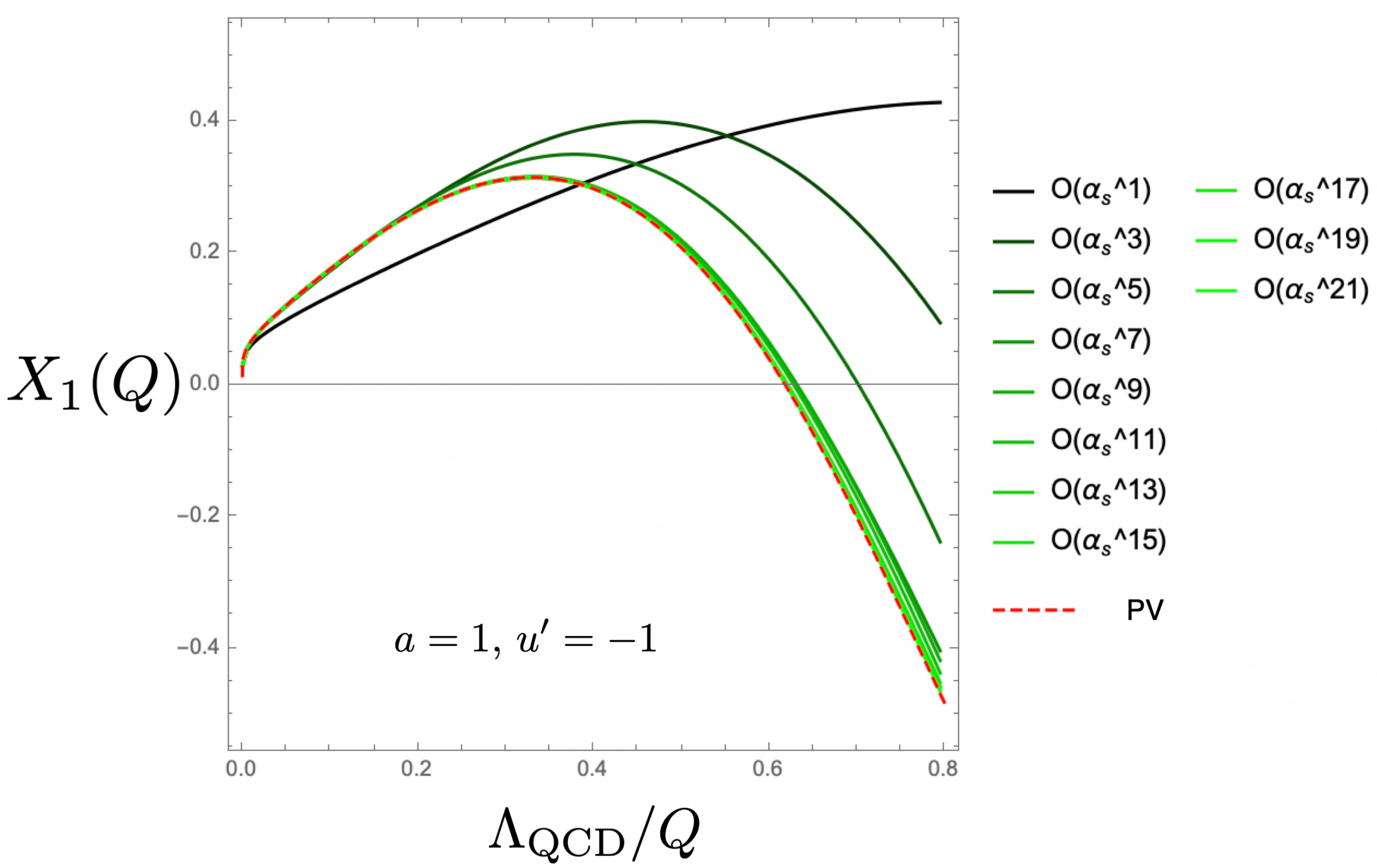}
\end{center}
\vspace*{-5mm}
\caption{\small Comparison of the PV prescription (dashed red line) 
and the FTRS method for $X_1(Q)$ (green lines with gradation).
We set $b_0=1$.
In the momentum space, we truncate the series at ${\cal O}(\alfs^{k+1})$
for $k=1,\cdots,21$.
}
\label{Fig:X1-analysis}
\vspace*{-4mm}
\end{figure}

\begin{figure}[t]
\begin{center}
\includegraphics[width=14cm]{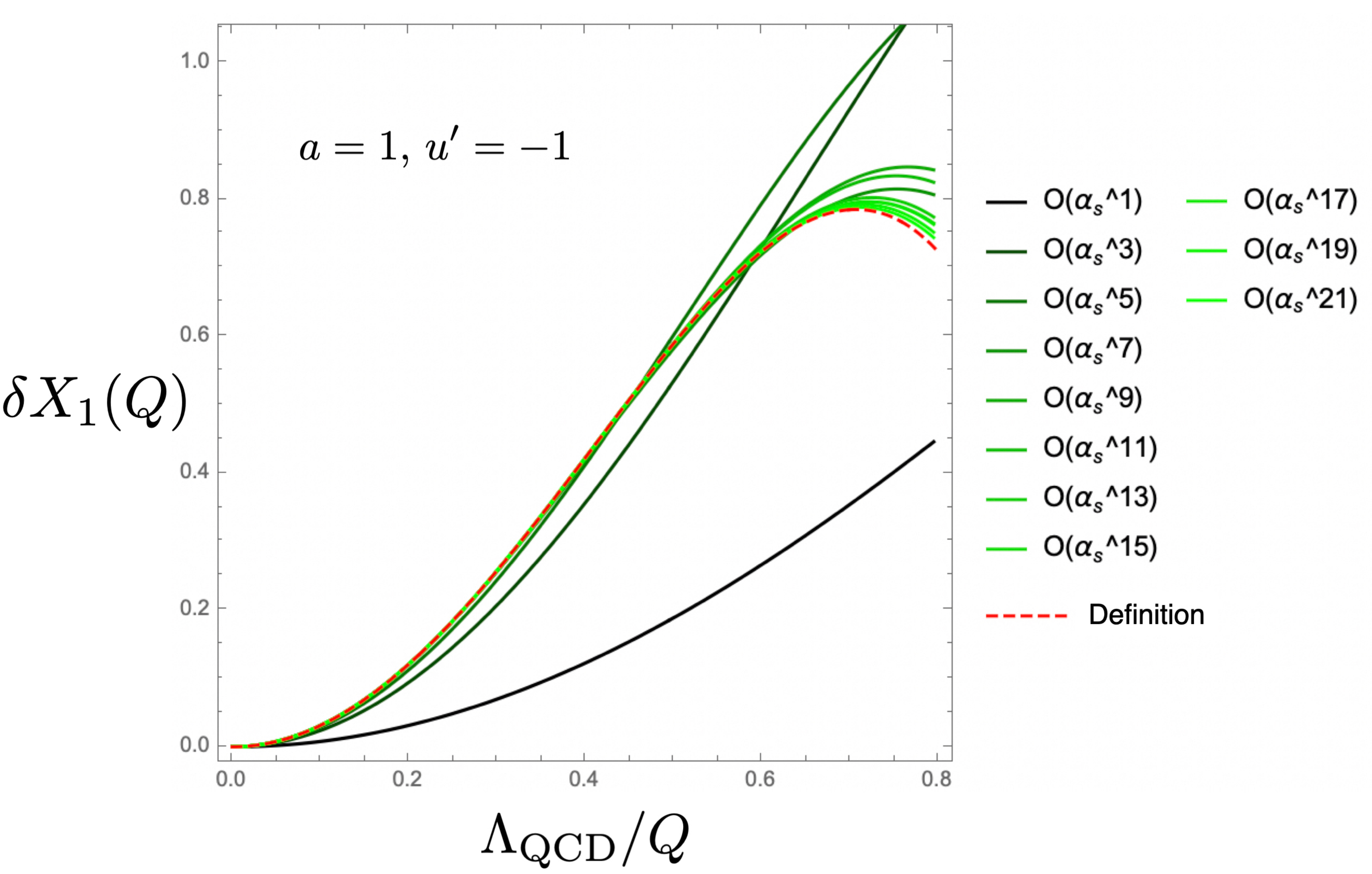}
\end{center}
\vspace*{-5mm}
\caption{\small Comparison of the behavior of the ambiguous part $\delta X_1(Q)$
calculated by the contour integral in the Borel plane [eq.~\eqref{deltaX}] (dashed red line) 
and the FTRS method (green lines with gradation).
We set $b_0=1$.
In the momentum space, we truncate the series at ${\cal O}(\alfs^{k+1})$
for $k=1,\cdots,21$.
}
\label{Fig:delta-X1-analysis}
\vspace*{-4mm}
\end{figure}

We compare the result of the FTRS method [eqs.~\eqref{X1FTRS}--\eqref{X1resum}] with
the (all-order) PV prescription result [eq.~\eqref{X1-PV}].
Fig.~\ref{Fig:X1-analysis} shows the result. 
The dashed red line represents the result of the PV prescription.
The green lines with gradation represent the result of the FTRS method
truncated at ${\cal O}(\alfs^{k+1})$ for $k=1,3,\cdots,21$.
As higher order terms are incorporated in the momentum-space series,
the result of the FTRS method approaches the PV-prescription result gradually.

In addition, we investigate the behavior of the ambiguous part.
The ambiguous part $\delta X$ is defined by
\bea
&&\big[\delta X_1(Q)\big]^{(k)}_{\rm FTRS}=\frac{1}{\pi Qi}\int_{C_*}\!\!d\tau\,\sin(\tau/Q)\sum_{n=0}^k \tilde{d}_n^{\rm subt}
\alfs(\tau)^{n+1}\non
&&~~~~~~
+\frac{1}{\pi Qi}\int_0^1dv\,\sinh(v)\int_{C_*}\!\!d\tau\,\sin(\tau v/Q)\sum_{n=0}^k \tilde{d}_n^{\rm resum}\alfs(\tau)^{n+1}\non
&&~~~~
=\frac{1}{\pi Qi}\int_{C_*}\!\!d\tau\,\sin(\tau/Q)\sum_{n=0}^k \tilde{d}_n^{\rm subt}
\alfs(\tau)^{n+1}\non
&&~~~~~~
+\frac{1}{\pi Qi}\int_{C_*}\!\!d\tau\,\frac{\cosh(1)\sin(\tau/Q)-\tau/Q\cos(\tau/Q)\sinh(1)}{1+\tau^2/Q^2}\sum_{n=0}^k \tilde{d}_n^{\rm resum}\alfs(\tau)^{n+1}.
\label{deltaX1-FTRS}
\eea
We compare eq.~\eqref{deltaX1-FTRS} to the desired behavior given by
\be
\delta X_1(Q)=\frac{\pi}{b_0}\Big(\LQ^2/Q^2-\LQ^4/Q^4\Big),
\label{deltaX1-PV}
\ee
which is obtained from the definition [eq.~\eqref{deltaX}].
The result is shown in Fig.~\ref{Fig:delta-X1-analysis}.
The dashed red line represents eq.~\eqref{deltaX1-PV}.
The green lines with gradation represent the result of the FTRS method
truncated at ${\cal O}(\alfs^{k+1})$ for $k=1,3,\cdots,21$.
We confirm that the result of the FTRS method gradually approaches the desired behavior
as high order terms are taken in.
It is non-trivial and interesting property of the FTRS method
that although the integrand of eq.~\eqref{deltaX1-FTRS} can be expanded as an infinite series of $\tau/Q$,
only the coefficients of the $\LQ^2/Q^2$ and $\LQ^4/Q^4$ terms 
(corresponding to IR renormalons) converge to non-zero value,
while those of the higher order terms of $\LQ/Q$ converge to zero, as the truncation order $k$ increases. 

We note that if IR renormalon ($u=u_{IR}$) remains in the momentum space,
the result of the FTRS method does not converge to that of the PV prescription.
In other words, there is always a possibility of an improper choice of the parameters $(a,u')$, 
in the case of a divergent series,
while any choice is allowed in the case of a convergent series.
The difference between the results of the FTRS method and the PV prescription is estimated 
to behave as $(\LQ/Q)^{2u_{IR}}$ by the closest uncanceled renormalon at $u=u_{IR}$.
It is discussed in more detail in Sec. 2.3 and App. G in ref.~\cite{Hayashi:2020ylq}.

\subsection{Model 3: IR and UV renormalons $(u=1,\,u=-1)$}
\label{sec3-3}
\subsubsection*{Resummation of ``original'' UV renormalons}
We can extend the formula for the resummation of the artificial UV renormalons in Sec.~\ref{sec2},
to the resummation of the UV renormalons originally contained in the perturbative series.
It is assumed that $X(Q)$ contains the UV renormalon at $u=-z$ ($z>0$),
whose singularity structure is a single pole.
We modify eq.~\eqref{tildeX-PT} as follows,
\be
\tilde{X}(\tau)=\frac{4\pi}{\tau^{3+2au'}}\bar{F}(\hat{H};z)\,(z+\hat{H})
\sum_{n=0}^\infty c_n\alfs(\tau^a)^{n+1}\,.
\ee
Here we define the function $\bar{F}(u;z)$
using $F(u)$ [eq.~\eqref{F(u)}],
\be
\bar{F}(u;z)=\frac{F(u)}{z+u}\,.
\ee
The factor $(z+\hat{H})$ eliminates the leading contribution from the UV renormalon at $u=-z$.
When the beta function is at one loop,
it can be seen as follows,
\bea
(z+\hat{H})\sum_{n=0}^\infty c_n\alfs(\tau^a)^{n+1}&&
=\sum_{n=0}^\infty \big(zc_n-(n+1)c_n\frac{\beta(\alfs)}{\alfs}\big)\alfs(\tau^a)^{n+1}\non
&&=\sum_{n=0}^\infty \big(zc_n+(n+1)c_nb_0\alfs\big)\alfs(\tau^a)^{n+1}\non
&&=zc_0\alfs(\tau^a)+\sum_{n=1}^\infty \big(zc_n+nb_0c_{n-1}\big)\alfs(\tau^a)^{n+1}\,,
\label{resumUV}
\eea
We approximate $c_n\approx N_{-z}(-b_0/z)^nn!$
at each term of the series in eq.~\eqref{resumUV},
then it is calculated as 
\be
zc_n+nb_0c_{n-1}\approx
N_{-z}\Big(z(-b_0/z)^{n}n!+nb_0(-b_0/z)^{n-1}(n-1)!\Big)\alfs(\tau^a)^{n+1}=0\,,
\label{UV-cancel}
\ee
which means that the sign-alternating divergent behavior of $c_n$ induced by $u=-z$ renormalon is canceled at each order of the series.
$\bar{F}(u;z)$ contains the UV renormalon pole at $u=-z$
as well as the UV renormalons generated by the Fourier transform.
To resum all the UV renormalons in $\bar{F}(u;z)$,
we separate the singular poles due to the UV renormalons from $\bar{F}(u;z)$,
and absorb the contributions of UV renormalons into the running coupling constant
using RGE in the way which is explained in Secs.~\ref{sec2}.
The explicit formula is given in the next section to avoid the cumbersome expression.
We note that this procedure does not request the information of the normalization of the UV renormalons\footnote{
In ref.~\cite{Lee:1999ws},
it is discussed how the normalization constant of the UV renormalons is estimated.
The normalization is calculated using the ordinary perturbative expansion of the Wilson coefficient
and the nature of the renormalon singularity in the Borel plane.}.
This is because the sign-alternating divergent behavior in the momentum-space series
is canceled exactly without knowing the normalization of the UV renormalon as in eq.~\eqref{UV-cancel},
and it is replaced to the RGE operation $1/(z+\hat{H})$
which can be converted to the resummation formula.

In this paper, the singularity structure of UV renormalons is assumed to be a simple pole.
The extension to the case with more generic structure of UV renormalons is left to future work.

\subsubsection*{Demonstration of resummation and subtraction of renormalons}
To perform the resummation of the UV renormalon and subtraction of the IR renormalon explicitly, 
we consider the simplified model $X_2$.
Its Borel transform is given by
\be
B_{X_2}(u)=\frac{e^{L_Qu}}{(1-u)(1+u)}=\sum_{n=0}^\infty\frac{f_n(L_Q)}{n!}\Big(\frac{u}{b_0}\Big)^n.
\ee
$B_{X_2}$ is singular at $u=1$ and $u=-1$ in the complex Borel $u$-plane.
The divergent perturbative series of $X_2$ is given by
\be
\big[X_2(Q)\big]_{\rm PT}=\sum_{n=0}^\infty f_n(L_Q)\alpha_s^{n+1}
=\sum_{n=0}^\infty f_n\alpha_s(Q)^{n+1},
\ee
where the coefficient $f_n=f_n(0)$ is explicitly given by
\be
f_n=b_0^nn!\frac{1+(-1)^n}{2}.
\ee 

Using the PV prescription, an exact prediction of $X_2$ from $B_{X_2}$ is calculated as
\bea
\big[X_2(Q)\big]_{\rm PV}
&=&\frac{e^{\frac{1}{b_0\alfs(Q)}}}{2b_0}\Gamma\Big(0,\frac{1}{b_0\alpha_s(Q)}\Big)+\frac{e^{-\frac{1}{b_0\alfs(Q)}}}{2b_0}{\rm Ei}\Big(\frac{1}{b_0\alfs(Q)}\Big).
\label{X2PV}
\eea
$\Gamma(a,z)$ is the incomplete gamma function defined by
\be
\Gamma(a,z)=\int_z^\infty dt \,t^{a-1}e^{-t}.
\ee

The Fourier transform to suppress the $u=1$ renormalon is defined by 
setting the parameters as $a=1$ and $u'=-1$.
The Fourier transform of $X_2$ is given by
\bea
\tilde{X}_2(\tau)&=&\frac{4\pi}{\tau}\bar{G}(\hat{H})(1+\hat{H})\sum_{n=0}^\infty f_n\alpha_s(\tau)^{n+1}\non
&=&\frac{4\pi}{\tau}\bar{G}(\hat{H})\sum_{n=0}^\infty n!b_0^n\alpha_s(\tau)^{n+1}.
\eea
$\bar{G}(u)$ is defined using the function $G(u)=\sin(\pi u)\Gamma(2u)$ [eq.~\eqref{tildeas}],
\be
\bar{G}(u)=\frac{G(u)}{1+u}=\frac{\sin(\pi u)\Gamma(2u)}{1+u}.
\ee
The UV renormalons in $\bar{G}(u)$ should be resummed.
We divide $\bar{G}(u)$ into two parts,
\be
\bar{G}(u)=\bar{G}^{\rm subt}(u)+\bar{G}^{\rm UV}(u)\,,
\ee
where
\be
\bar{G}^{\rm subt}(u)=\frac{\sin(\pi u)}{1+u}\bigg[\Gamma(2u)-\frac{1}{4(1+u)}+\sum_{\ell=0}^\infty\frac{1}{(2\ell+1)!}\frac{1}{2u+2\ell+1}\bigg]\,,
\ee
and
\be
\bar{G}^{\rm UV}(u)=\frac{\sin(\pi u)}{1+u}\frac{1}{4(1+u)}-\frac{\sin(\pi u)}{1+u}\sum_{\ell=0}^\infty\frac{1}{(2\ell+1)!}\frac{1}{2u+2\ell+1}\,.
\ee 
The formula for the resummation of the generated UV renormalons is given by
\be
\tilde{X}_2(\tau)=\big[\tilde{X}_2(\tau)\big]_{\rm subt}+\big[\tilde{X}_2(\tau)]_{\rm resum}.
\ee
The first series, which is not affected by the IR and UV renormalons, is given by
\be
\big[\tilde{X}_2(\tau)]_{\rm subt}
=\frac{4\pi}{\tau}\sum_{n=0}^\infty \tilde{f}_n^{\rm subt}\alpha_s(\tau)^{n+1},
\ee
where $\tilde{f}_n^{\rm subt}$ is determined from 
\bea
&&\sum_{n=0}^\infty \tilde{f}_n^{\rm subt}\alpha_s^{n+1}
=\bar{G}^{\rm subt}(\hat{H})\sum_{n=0}^\infty n!\alpha_s^{n+1}\non
&\approx&\frac{1}{b_0}\Big[
1.57 x + 0.723 x^2 - 0.830 x^3 - 0.0625x^4 + 
 3.61x^5 - 2.29x^6 - 29.5x^7+\cdots\Big]\,,
\eea
where $x=b_0\alfs$.
The other series after the resummation of the UV renormlaons is given by
\bea
\big[\tilde{X}_2(\tau)]_{\rm resum}
=\frac{4\pi}{\tau}\int_0^1dv\Big(\frac{\sinh(v)}{v}-\frac{v}{2}\Big)\sum_{n=0}^\infty \tilde{f}_n^{\rm resum}\alpha_s(\tau/v)^{n+1},
\eea
with $\tilde{f}_n^{\rm resum}$ is determined from 
\bea
\sum_{n=0}^\infty \tilde{f}_n^{\rm resum}\alpha_s^{n+1}
&=&-\frac{\sin(\pi \hat{H})}{1+\hat{H}}\sum_{m=0}^\infty m!\alpha_s^{m+1}\non
&=&\sum_{k,\ell,m=0}^\infty\frac{(-1)^{k+\ell+1}}{(2\ell+1)!}(2\ell+k+m+1)!\pi^{2\ell+1}\,\alpha_s^{2\ell+k+m+2}\non
&\approx&\frac{1}{b_0}\Big[-3.14x^2 + 12.2x^4 - 
 62.9x^6 + 379x^8+\cdots\Big]\,,
\eea
where $x=b_0\alpha_s$.


The one-parameter integral form of $\big[X_2(Q)\big]_{\rm FTRS}$
in the case of the truncation at ${\cal O}(\alfs^{k+1})$
is given by

\be
\big[X_2(Q)\big]_{\rm FTRS}^{(k)}=\big[X_2(Q)\big]_{\rm subt}^{(k)}+\big[X_2(Q)\big]_{\rm resum}^{(k)}\,,
\label{X2FTRS}
\ee
where each part is given by
\bea
&&\big[X_2(Q)\big]^{(k)}_{\rm subt}
=\frac{2}{\pi Q}\int_{0}^\infty \!\!dt\,\exp(-t/Q)\sum_{n=0}^k \tilde{f}_n^{\rm subt}
\,{\rm Re}\big[\alfs(it)^{n+1}\big]\non
&&~~~~~~~~~~~~~~~~~
+\frac{1}{\pi Qi}\int_{C_*}\!\!d\tau\,\cos(\tau/Q)\sum_{n=0}^k \tilde{f}_n^{\rm subt}
\alfs(\tau)^{n+1}\,,
\label{X2subt}
\eea
and
\bea
&&\big[X_2(Q)\big]^{(k)}_{\rm resum}
=\frac{2}{\pi Q}\int_0^1dv\,\Big(\sinh(v)-\frac{v^2}{2}\Big)\int_{0}^\infty \!\!dt\,\exp(-tv/Q)\sum_{n=0}^k 
\tilde{f}_n^{\rm resum}
\,{\rm Re}\big[\alfs(it)^{n+1}\big]\non
&&~~~~~~~~~~~~~~~~~
+\frac{1}{\pi Qi}\int_0^1dv\,\Big(\sinh(v)-\frac{v^2}{2}\Big)\int_{C_*}\!\!d\tau\,\cos(\tau v/Q)\sum_{n=0}^k \tilde{f}_n^{\rm resum}\alfs(\tau)^{n+1}\non
&&=\frac{2}{\pi Q}\int_{0}^\infty \!\!dt\,\Bigg\{\frac{-2+e^{-t/Q} \left(t^2/Q^2+2 t/Q+2\right)}{2 t^3/Q^3}\non
&&~~~~~~~~~~~~~~~~~~~~~
+\frac{1-e^{t/Q}\big(t/Q \sinh (1)+\cosh
   (1)\big)}{t^2/Q^2-1}\Bigg\}\sum_{n=0}^k \tilde{f}_n^{\rm resum}
\,{\rm Re}\big[\alfs(it)^{n+1}\big]\non
&&
+\frac{1}{\pi Qi}\int_{C_*}\!\!d\tau\,\Bigg\{\frac{\left(-\tau^2/Q^2+2\right) \sin (\tau/Q)-2 \tau/Q \cos (\tau/Q)}{2 \tau^3/Q^3}\non
&&~~~~~~~~~~~~~~~~~~~~~
+\frac{\tau/Q \sinh (1) \sin (\tau/Q)+\cosh
   (1) \cos (\tau/Q)-1}{\tau^2/Q^2+1}\Bigg\}\sum_{n=0}^k \tilde{f}_n^{\rm resum}\alfs(\tau)^{n+1}\,.\non
\label{X2resum}
\eea

\begin{figure}[t]
\begin{center}
\includegraphics[width=14cm]{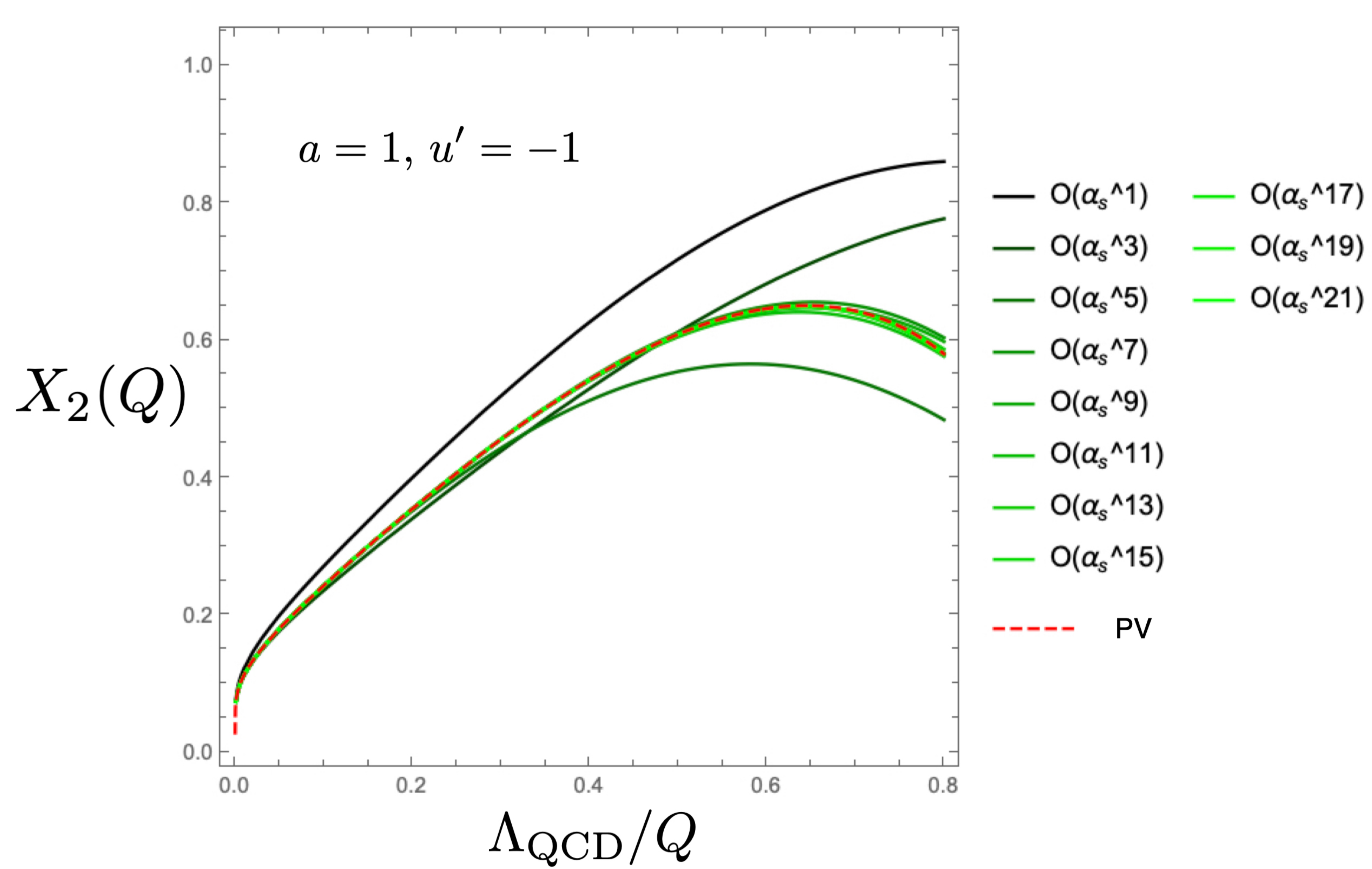}
\end{center}
\vspace*{-5mm}
\caption{\small Comparison of the PV prescription (dashed red line) 
and the FTRS method for $X_2(Q)$ (green lines with gradation).
We set $b_0=1$.
In the momentum space, we truncate the series at ${\cal O}(\alfs^{k+1})$
for $k=1,\cdots,21$.
}
\label{Fig:X2-analysis}
\vspace*{-4mm}
\end{figure}

We compare the result of the FTRS method [eqs.~\eqref{X2FTRS} -- \eqref{X2resum}] 
with the (all-order) PV
prescription result [eq.~\eqref{X2PV}]. 
Fig.~\ref{Fig:X2-analysis} shows the result. The dashed red line
represents the result of the PV prescription. The green lines with gradation represent
the result of the FTRS method truncated at ${\cal O}(\alfs^{k+1})$ 
for $k=1,3,\cdots,21$. 
As higher order terms are incorporated in the momentum series, 
the result of the FTRS method approaches the PV-prescription result gradually.

\section{Conclusion and discussions}
\label{sec4}
In order to remove the renormalons in the leading Wilson coefficient in the framework of OPE,
it is necessary to separate the ambiguous contribution generated by the IR renormalons. 
The FTRS method, which was proposed as a new renormalon-subtraction method, 
is applied to several toy models with the QCD beta function at one loop, 
and we compared the results with those of the PV prescription to 
investigate quantitatively how this method works.

One practical improvement over refs.~\cite{Hayashi:2020ylq,Hayashi:2021vdq} is that 
we have resummed all the UV renormalons that appear in the momentum space. 
Thanks to this improvement, the perturbation series in the momentum space 
is completely unaffected by the artificial UV renormalons,
while keeping the suppression of IR renormalons by the sine factor.
Thus the effect of renormalons on the results of the FTRS method is purely the effect of them included in the original series.

First of all, as an example of applying the FTRS method to a convergent series, 
we considered the running coupling $\alfs(Q)$.
We examined how the result of the FTRS method ($a=1,\,u'=-1$) 
with the truncated momentum-space series, 
behaves depending on the truncation order. 
The result shows that as the truncation order is increased, 
the result gradually approaches that of the PV prescription 
(an exact expression of the one-loop running coupling). 
This implies that the convergent series returns to the original series 
even after the FTRS method is applied. 
The choice of the parameters $(a,\,u')$ for the Fourier transform should be arbitrary 
for series that do not contain renormalons. 
As an example, we adjust $a=1/2,\, u'=-1$ 
and then confirm 
the convergence of the result of the FTRS method to that of the PV prescription
(Fig.~\ref{Fig:as-analysis-b}). 

Next, we applied the FTRS method ($a=1,\,u'=-1$) to the model $X_1(Q)$ 
with IR renormalions at $u=1,\,2$ in order to show that the multiple renormalons can be subtracted simultaneously, which is characteristic of the FTRS method. 
The result of the FTRS method gradually approaches that of the PV prescription 
as the truncation order in the momentum space is increased.
Also, for the ambiguous part $\delta X_1$ separated by the FTRS method, 
we found the result of the FTRS method approaches the behavior defined by eq.~\eqref{deltaX} using the Borel transform. 
This indicates that our method is also useful for estimating the normalization of the IR renormalons.

Finally, for the series $X_2(Q)$ which originally contains both IR and UV renormalons, 
we devised a method to resum the contribution of the UV renormalon by applying the FTRS method, and compared the results with those of the PV prescription. 
In the case of UV renomalons of pole-singularity type, 
the method proposed in Sec.~\ref{sec3-3} can resum the contribution of them 
without knowing their normalization. 
The result of the FTRS method gradually approaches that of the PV prescription. 
Recently, it has been discussed
that the pole-$\msbar$ relation for quark mass
is affected by the UV renormalon at $u=-1$ 
and the effect is possibly visible in the latest analysis \cite{Ayala:2019hkn}.
In the large-$\beta_0$ approximation, the type of the singularity
of the pole-$\msbar$ relation in the Borel plane 
is a simple pole \cite{Beneke:1994qe,Hayashi:2019mlb}.
Our new procedure will be useful to resum the leading contribution of it.

Resummation prescriptions for a general UV renormalon structure are left to a future work. 
Analysis using OPE and RGE can determine the structure of the singularities of UV renormalons in the Borel plane,
which is almost same as the structure of IR renormalons \cite{Beneke:1998ui,Lee:1999ws}.
In a previous paper \cite{Hayashi:2021vdq}, we have shown how to suppress a general form of IR renomalons in momentum space,
which would be a hint for the case of UV renormalons.

\vspace{1cm}

The author would like to thank to Yukinari Sumino and Hiromasa Takaura 
for fruitful discussions.
He acknowledges support from GP-PU at Tohoku University.
The work was also supported in part by Grant-in-Aid for 
JSPS Fellows (No. 21J10226) from MEXT, Japan.

\begingroup\raggedright\endgroup

\end{document}